\documentclass[useAMS,usenatbib]{mn2e}
\usepackage{psfig, epsf, epsfig}
\title[The origin of the Magellanic Clouds]{When was the Large Magellanic Cloud
accreted onto the Galaxy ?}
\author[K. Bekki]
{Kenji Bekki${}^1$\thanks{E-mail:
bekki@cyllene.uwa.edu.au}\\
       ${}^1$ICRAR M468
The University of Western Australia
35 Stirling Hwy, Crawley
Western Australia, 6009 \\}

\begin{document}

\date{Accepted, Received 2005 February 20; in original form }

\pagerange{\pageref{firstpage}--\pageref{lastpage}} \pubyear{2005}

\maketitle

\label{firstpage}

\begin{abstract}
Using fully self-consistent N-body models for the dynamical evolution
of the Large Magellanic Cloud (LMC)  in the Galaxy,
we show that if the LMC initially has an extended old  stellar halo
before its commencement of tidal interaction with the Galaxy, 
physical properties of the stars stripped from the LMC stellar halo
can have fossil information as to when and where
the LMC was accreted onto the Galaxy for the first time.  
If the epoch of the first LMC accretion onto the Galaxy 
from outside its viriral radius is more than $\sim 4$  Gyr ago
(i.e., at least two pericenter passages), 
the stars stripped from the stellar halo of the LMC can form an irregular
polar ring or a thick  disk
with a size of $\sim 100$ kpc and rotational kinematics.
On the other hand, if the LMC  was first accreted onto
the Galaxy quite recently ($\sim 2$ Gyr
ago),
the stripped stars form shorter leading and trailing stellar stream at $R=50-120$ kpc.
Also distributions of the  stripped stars 
in phase space between the two cases can be significantly different.
The derived  differences in structure and kinematics of the stripped stars
therefore  suggest that if we compare the observed three-dimensional (3D)  distribution
and kinematics of the
outer Galactic stellar halo along the polar-axis,
then we can give strong constraints on the past orbit of the LMC. 
 We also discuss whether
the orbital properties of the LMC in successful formation models for the Magellanic
Stream (MS) can be consistent with
orbital properties of the LMC-type systems in Galaxy-type halos predicted from
recent high-resolution cosmological simulations in a $\Lambda$CDM universe.
We find that the orbital properties of the LMC in the successful formation models
are consistent with those predicted from the cosmological simulations. 
We also find that the LMC can not merge with the Galaxy within the last $\sim 6$ Gyr
in models consistent with predictions from  the $\Lambda$CDM simulations.
Given that the successful MS formation 
models predict at least two pericenter passages of the LMC
in the Galaxy,
we conclude that the LMC was accreted onto the Galaxy more than  $\sim 4$ Gyr ago
so that interaction between the LMC, the Small Magellanic Cloud (SMC),
and the Galaxy could form the MS
and its Leading Arms (LAs).
\end{abstract}


\section{Introduction}

Recent proper motion measurements of the LMC and the Small Magellanic Cloud
(SMC) have suggested that the LMC is now  moving around the Galaxy with
a velocity ($V_{\rm LMC}$) of $\sim 380$ km s$^{-1}$ with respect to the Galaxy 
(e.g., Kallivayalil et al. 2006, K06;  Piatek et al. 2008). The latest observational
studies on the proper motions of the Clouds (Vieira et al. 2010, V10) 
have however derived significantly
different $V_{\rm LMC}$ ($=343 \pm47.8$ km s$^{-1}$), which is significantly
smaller than $V_{\rm LMC}$ derived by the above earlier studies (see also Costa
et al. 2009, C09, which derived $V_{\rm LMC} \sim 300$ km s$^{-1}$).
Using N-body models of the LMC,
Bekki (2011) has recently shown that if the total number of 
the LMC fields used for the derivation
of the center-of-mass proper motion of the LMC, 
then the derived center-of-mass
proper motion of the LMC can significantly deviate from the true one. 
These observational and theoretical studies imply that
the proper motion of the LMC observationally derived so far does not allow
us to construct a model for the very precise 3D orbit of the LMC. 

\begin{table*}
\centering
\begin{minipage}{175mm}
\caption{Description of the  model parameters for
the representative models.}
\begin{tabular}{cccccccccc}
{Model}
& {$M_{\rm dm, mw}$
\footnote{The total dark matter halo  mass of the Galaxy  in units of ${\rm M}_{\odot}$.}}
& {$r_{\rm vir, mw}$
\footnote{The virial radius of the Galactic dark matter halo  in units of kpc.}}
& {$c$
\footnote{The $c$ parameter of the Galactic dark matter halo.}}
& {$f_{\rm v}$
\footnote{The ratio of the velocity of the LMC ($v_{\rm L}$)
to the circular velocity  ($v_{\rm cir}$)
at the initial position of the LMC .}}
& {$f_{\rm v, y}$
\footnote{The initial tangential  velocity of the LMC is given as $f_{\rm v,y}v_{\rm cir}$.}}
& {$f_{\rm v, z}$
\footnote{The initial radial  velocity of the LMC is given as $f_{\rm v,z}v_{\rm cir}$.}}
& { $R_{\rm h, L}$
\footnote{The initial size of the stellar halo in the LMC in units of $R_{\rm d, L}$,
which is the size of the LMC stellar disk.}} 
& {SMC 
\footnote{Presence (``Yes'') or absence (``No'') of the SMC in the model.}}
& {comments
\footnote{The initial distance between the LMC and the SMC is represented by
$R_{\rm i, S}$ for the models with SMC.}} \\
M1 & $10^{12}$  & 245 & 10 & 0.5 & 0.5 & 0.0 & 3 & No & the standard FPS \\
M2 & $10^{12}$  & 245 & 10 & 0.6 & 0.6 & 0.0 & 3 & No & the standard SPS \\
M3 & $10^{12}$  & 245 & 10 & 1.1 & 0.5 & 1.0 & 3 & No &  \\
M4 & $10^{12}$  & 245 & 10 & 1.1 & 0.6 & 0.9 & 3 & No &  \\
M5 & $10^{12}$  & 245 & 10 & 0.6 & 0.5 & 0.4 & 3 & No &  \\
M6 & $10^{12}$  & 245 & 10 & 0.6 & 0.6 & 0.0 & 2 & No &  smaller size of the LMC halo\\
M7 & $10^{12}$  & 245 & 10 & 0.6 & 0.6 & 0.0 & 3 & Yes & $R_{\rm i, S}=8R_{\rm d, L}$ \\
M8 & $10^{12}$  & 245 & 10 & 0.6 & 0.6 & 0.0 & 3 & Yes & $R_{\rm i, S}=6R_{\rm d, L}$ \\
M9 & $2\times 10^{12}$  & 347 & 9 & 0.5 & 0.0 & 0.0 & 3 & No &  more massive Galactic halo\\
M10 & $2\times 10^{12}$  & 347 & 9 & 1.1 & 0.4 & 1.0 & 3 & No &  \\
M11 & $2\times 10^{12}$  & 347 & 9 & 1.1 & 0.6 & 0.9 & 3 & No &  \\

\end{tabular}
\end{minipage}
\end{table*}

The past 3D motions of the LMC and the SMC are the most important parameters 
in the construction of the formation model for MS and LAs
(Murai \& Fujimoto 1980). A recent ``first passage'' scenario has shown
that MS was formed owing to LMC-SMC interaction well before the accretion of the Clouds
onto the Galaxy for the first time (Besla et al. 2010). Although this model
has failed to explain the observed bifurcated structure of the MS and the elongated
leading arm,  it is consistent with some of the observational results of  the above-mentioned
HST proper
motions of the  Clouds.  On the other hand,  classical tidal models can better
explain physical properties of MS and LA (e.g., Gardiner \& Noguchi 1996, GN96;
Yoshizawa \& Noughci 2003, YN03; Connors et al. 2006, C06),
their models appear to be inconsistent with the above-mentioned HST
proper motion results.
Given that there are significant differences in orbital evolution of the Clouds
between these previous models,  it is observationally important
to give more stringent constraints on the 3D motions of the Clouds, in particular,
the LMC which gravitationally dominates the Magellanic system.

Recently photometric and spectroscopic observations of stars in the surrounding regions
of the LMC have revealed possible evidence for the presence of 
an extended  stellar halo of the LMC  (e.g., Minniti et al. 2003; Mu\~noz et al. 2006;
Majewski et al. 2009; Saha et al. 2010). 
For example, Minniti et al. (2003) found that the velocity dispersion of the 43 RR Lyrae
stars in the LMC region is $53 \pm 10$ km 2$^{-1}$ and suggested that the LMC has a 
dynamically hot stellar halo.  Mu\~noz et al. (2008) and Majewski et al.  (2009) 
have revealed a possible presence of the LMC stars out to $22-23^{\circ}$ 
from the LMC center
(corresponding roughly 10 times the scale radius of the LMC stellar disk),
which can possibly consist of the outer halo of the LMC.
Subramaniam \&  Subramaniam (2009) found that the distribution of RR Lyrae stars
in the inner region of the LMC has a disk-like distribution and suggested
that the old RR Lyrae stars do not trace the extended metal-poor stellar halo
of the LMC.

The possible extended stellar halo of the LMC can be much more strongly influenced by
the Galaxy than the stellar disk 
during the LMC-Galaxy tidal interaction.
Therefore the stellar halo
can be stripped much more efficiently
by the Galaxy to become a part of the Galactic stellar halo.
Thus, the outer part of the Galactic stellar halo can contain stars initially
from the stellar halo of the LMC and therefore can have fossil records on
the past orbital history of the LMC. 
However,
the tidal stripping of the {\it original metal-poor} stellar halo of the LMC 
has not been investigated so far in previous numerical simulations.
It is thus unclear what structural and kinematical properties
of the stripped halo stars of the LMC have
in the outer stellar halo of the Galaxy.

The purpose of this paper is thus to propose that if the LMC  
has an extended stellar halo before its commencement of the LMC-Galaxy
tidal interaction,  then the stripped halo stars in the outer Galactic halo
can have fossil information as to when and 
where the LMC was first accreted onto the Galaxy. 
Using self-consistent N-body simulations of the LMC-Galaxy tidal interaction, 
we investigate how the tidal field of the Galaxy can influence
the dynamical evolution of the stellar halo of the LMC.
We show how the 3D spatial distribution and kinematics of stars
stripped from the LMC stellar halo  depend on the past 3D orbits
of the LMC with respect to the Galaxy.

\section{The model}

We investigate how the LMC dynamically evolves after it is first accreted onto the Galaxy
from outside the virial radius of the dark matter halo of the Galaxy.
In the present model, both the LMC and the Galaxy are represented by N-body particles
so that dynamical friction of the LMC against the Galactic dark matter halo
can be self-consistently modeled.
The present model is therefore  more sophisticated than those used in recent 
models for the Magellanic system
by Besla et al. (2010) and Diaz \& Bekki (2011, DB11) which treated the Galaxy as a fixed
gravitational field. 
Owing to the smaller
mass of the SMC in comparison with the LMC mass,  the SMC does not significantly
influence the orbital evolution of the LMC.
Also we focus exclusively on how  the stellar halo initially in the LMC
dynamically evolves during the LMC-Galaxy tidal interaction.
Thus we do not investigate extensively
the models with  the SMC in the present study for clarity.
We however briefly discuss whether the SMC can influence the evolution of the LMC
stellar halo using two representative models with the SMC.

The LMC is modeled as a bulge-less disk galaxy embedded in a massive dark matter halo.
Since the LMC is assumed to fall onto the Galaxy from outside the dark matter halo of 
the Galaxy,  it is highly likely that the LMC initially 
has an  extended dark matter halo
(i.e., not truncated at the tidal radius of the LMC).
The total  mass and the virial radius  of the dark matter halo  of the LMC 
are $M_{\rm dm, L}$ and $r_{\rm vir, L}$, respectively.
We adopted an NFW halo density distribution (Navarro, Frenk \& White 1996)
suggested from CDM simulations:
\begin{equation}
{\rho}(r)=\frac{\rho_{0}}{(r/r_{\rm s})(1+r/r_{\rm s})^2},
\end{equation}
where  $r$, $\rho_{0}$, and $r_{\rm s}$ are
the spherical radius,  the characteristic  density of a dark halo,  and the
scale
length of the halo, respectively.
We adopted $c=12$ ($=r_{\rm vir, L}/R_{\rm s}$)
and
the mass ratio of $M_{\rm dm, L}$ to $M_{\rm d, L}$
was fixed at 16.7.

The stellar disk with the size $R_{\rm d, L}$
and mass $M_{\rm d, L}$ is assumed to have an exponential
profile with the scale length ($a_{\rm d, L}$) of 1.5 kpc and the truncation radius
of 7.5 kpc ($=R_{\rm d, L}$). In order to investigate the models
which have the maximum rotation speed of $\sim 120$ km s$^{-1}$
as suggested by recent observations by Piatek et al. 2008, 
we assume that the total mass of the LMC ($M_{\rm t, L}$)
is $9.6 \times 10^{10} {\rm M}_{\odot}$ and 
$r_{\rm vir, L}/R_{\rm d, L}=10.0$.  
The LMC has a stellar halo with the mass $M_{\rm h, L}$ and
the size $R_{\rm h, L}$ and 
the radial density profile is assumed to follow the power-law one
with the power-law index of $-3.5$  (${\rho}(r) \sim r^{-3.5}$).
We mainly investigate the models with $M_{\rm h, L}/M_{\rm d, L}=0.01$
and $R_{\rm h, L}=3R_{\rm d, L}$.
Gas dynamics, star formation, and chemical evolution modeled in our
previous models (Bekki \& Chiba 2005, 2009) are not included in the present
model, because we focus exclusively on the dynamical evolution of the stellar halo
of the LMC.

Observational studies for 1047 edge-on disk galaxies in the Sloan Digital 
Sky Survey (SDSS) revealed almost ubiquitous presence of stellar halos around
disk galaxies (Zibetti et al. 2004). Their Fig. 2 clearly shows that the stellar halos
extend at least $10a_{\rm disk}$, where $a_{\rm disk}$ is the scale length of a 
stellar disk.  The Galaxy is observed to have the stellar halo which extends at least
30 kpc corresponding to $\sim 9a_{\rm disk}$ 
($\sim 12a_{\rm disk}$) for the stellar scale length of 3.5 (2.5) kpc
(e.g., Chiba \& Beers 2000) and has a mass that is about $\sim 1$\% of the Galactic
stellar disk (Freeman \& Bland-Hawthorn 2002).
Thus the present models with $M_{\rm h, L}/M_{\rm d, L}=0.01$
and $R_{\rm h, L}=3R_{\rm d, L}$ ($=15a_{\rm d, L}$) 
are quite reasonable.
We however investigate models with different  $R_{\rm h, L}$ in order to make more 
robust conclusions of the present numerical simulations.

The Galaxy is assumed to consist of the NFW dark matter halo,  exponential stellar disk,
and Hernquest bulge. The dark halo has a mass $M_{\rm dm, mw}$,
a virial radius $r_{\rm vir, mw}$,  and a $c$ parameter.
We investigate two dark halo  models 
in which the maximum circular velocity can be $\sim 220$
km s$^{-1}$ that is observed for the Galaxy. One has   
$M_{\rm dm, mw} =10^{12} {\rm M}_{\odot}$,
$r_{\rm vir, mw}=$245 kpc, and $c=10$
and the other has
$M_{\rm dm, mw} =2 \times 10^{12} {\rm M}_{\odot}$,
$r_{\rm vir, mw}=$347kpc, and $c=9$.
The stellar disk has a mass of $6 \times 10^{10} {\rm M}_{\odot}$,
a size of 17.5 kpc (scale-length of 3.5 kpc), and the Q-parameter of 1.5.
The bulge has a mass of $10^{10} {\rm M}_{\odot}$ and a size of 3.5 kpc 
(scale-length of 700pc). Like the LMC model, the Galaxy is assumed to have no gas
and no star formation. The disk plane of the Galaxy is set to be the $x$-$y$ plane
in the present study (i.e., the $z$-axis is the polar direction of the Galaxy).

The LMC is initially located at $R_{\rm i}$ from the center of the Galaxy
and has a velocity ($v_{\rm L}$) of $f_{\rm v} v_{\rm c}$,
where $v_{\rm c}$ is the circular velocity at $R_{\rm i}$
and $f_{\rm v}$ is a parameter controlling the 3D orbit of the LMC.  
The initial orbital plane of the LMC is inclined by ${\Theta}_{\rm L}$ degrees with respect
to the Galactic plane (i.e., the $x$-$z$ plane).
$R_{\rm i}$ is fixed at $1.07r_{\rm vir, mw}$  for all models 
so that the LMC can be accreted from outside the virial radius of the Galaxy.
Previous best  models for the MS formation predicted that
the orbital plane of the LMC is almost coincident with the $y$-$z$ plane
(e.g., YN03).
We thus assume that ${\Theta}_{\rm L}=90$: the spin vector 
of the initial orbital angular momentum of the LMC is the same as the $x$-axis.
 The initial velocity components of the LMC  are  given
as (0, $f_{\rm v, y}v_{\rm c}$, -$f_{\rm v, z}v_{\rm c}$),
where ${f_{\rm v,y}}^2+{f_{\rm v,z}}^2={f_{\rm v}}^2$. The $y$- and $z$-components
of the velocity correspond to initial tangential and radial velocities, respectively.
The spin vector of the stellar disk of the LMC is inclined by 45 degrees with respect
to the orbital plane of the LMC for all models.

By changing $f_{\rm v}$ (also $f_{\rm v,y}$ and $f_{\rm v,z}$), 
we can determine the time (referred to as $T_{\rm p}$
from now on)
when (i) the LMC-Galaxy distance
can become the observed 50 kpc
and (ii) the LMC is moving away from the Galaxy.
We mainly investigate the dark matter model of the Galaxy with
$M_{\rm dm, mw}=10^{12} {\rm M}_{\odot}$.
If we adopt $f_{\rm v}=0.5$ for the above dark matter model, then
$T_{\rm p}$ is just after the first pericenter passage: this model corresponds
to the first passage scenario (referred to as ``FPS''  and  the model M1).
If we adopt $f_{\rm v}=0.6$, then $T_{\rm p}$ is just after two  pericenter passages:
this corresponds to the second pericenter passage scenario
(``SPS'' and M2). Fig. 1 shows the orbital evolution of the LMC in the standard FPS 
(M1) and SPS (M2).  
We mainly discuss the results of these standard FPS and SPS 
in order to more clearly show whether  stars stripped from the LMC stellar halo
have fossil information of the past orbital history of the LMC: we however investigate
different models and discuss the results briefly in \S 3.2.
The parameter values  for the representative 11 models investigated in the present study
are shown in the Table 1.

If we adopt models with larger $v_{\rm f}$ ($\ge 0.7$),
then $T_{\rm p}$ can be after the third or fourth pericenter passage of the LMC.
The simulated distribution of the stripped LMC halo stars in these models
are essentially
similar to  that in the SPS.
The aim of this paper is to more clearly show how the spatial distribution
of the stripped halo stars depends on the past orbit of the LMC. 
Accordingly we focus on the differences in physical properties   
the stripped halo stars   between the FPS and SPS 
in the present paper. 
Also it should be stressed that 
the present models can not reproduce exact locations of the LMC  
in the Galactic coordinate
(i.e., ($x$, $y$, $z$) for the LMC): we adopt the present models 
in order to point out the possible differences in physical properties of the stripped
LMC stellar halo between the two scenarios. Fully self-consistent 3D models
which can explain the present 3D locations and velocities
of the LMC and the SMC will be discussed in forthcoming papers. 
 
We estimate the velocity of the LMC when (i) the LMC passes the pericenter
around the Galaxy and (ii) the pericenter distance ($R_{\rm p}$)
is close to $\sim 50$kpc that corresponds
to the present distance of the LMC from the center of the Galaxy. 
We confirm that the velocity at $R_{\rm p}$ ($V_{\rm p}$) is in good agreement
with the observed one ($300-380$ km s$^{-1}$; K06; C09; V10) for all models
except the model M11 with $R_{\rm p}=91$ kpc (which is therefore unreasonable
for the LMC). For example,  $V_{\rm p}$ is 323 km s$^{-1}$, 300 km s$^{-1}$,
342 km s$^{-1}$, 347 km s$^{-1}$, and 355 km s$^{-1}$ for models M1, M2, M3, M4,
and M9, respectively. For all models,  the LMC can not merge with the Galaxy within
$\sim 6$ Gyr after the first passage of the Galactic virial radius. These results
demonstrate that the present LMC model is reasonable and realistic in discussing
dynamical evolution of the LMC.

Numerical computations
were carried out both on 
the latest version of GRAPE
(GRavity PipE, GRAPE-DR) -- which is the special-purpose
computer for gravitational dynamics (Sugimoto et al. 1990) at
University of Western Australia (UWA) and on the HP 390S blade  machine 
with three GPU cards (NVIDIA Tesla M2070) with  CUDA G5/G6 software package being
installed for calculations of gravitational dynamics at iVEC in UWA.
The total number of particles used for dark matter halo, stellar disk,
and bulge of the Galaxy in a simulation are 800000, 200000, and 33400, respectively.
The total number of particles used for dark matter, stellar halo, and stellar disk
in the LMC are 100000, 100000, and 50000, respectively. The adopted  gravitational softening
length is fixed at 200pc for all models.  
The time $T$  represents the time that has elapsed since
the simulation starts.  The polar axis of the Galactic disk
does not change during a simulation in the present study owing to the adopted LMC orbit.
The LMC disk particles  are not stripped owing to the adopted large initial mass
of the LMC ($M_{\rm t, L}=9.6 \times 10^{10} {\rm M}_{\odot}$).

\begin{figure}
\psfig{file=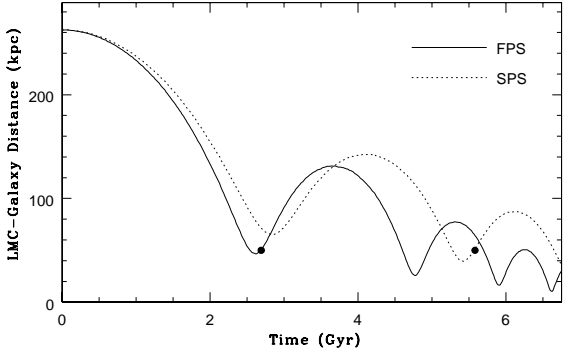,width=8.0cm}
\caption{
The time evolution of the LMC-Galaxy distance in the standard FPS (solid,
model M1) and SPS (dotted, M2). 
Filled circles represent the present LMC-Galaxy distances in these two models.
}
\label{Figure. 1}
\end{figure}

\begin{figure}
\psfig{file=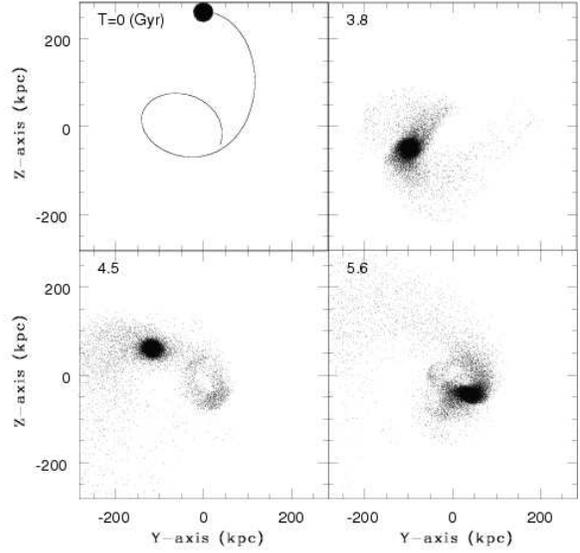,width=8.0cm}
\caption{
The time evolution of the distribution of the LMC stellar halo projected onto 
the $y$-$z$ plane in the SPS.
The solid line represent the past 5.6 Gyr orbit of the LMC in the SPS.
}
\label{Figure. 2}
\end{figure}

\begin{figure}
\psfig{file=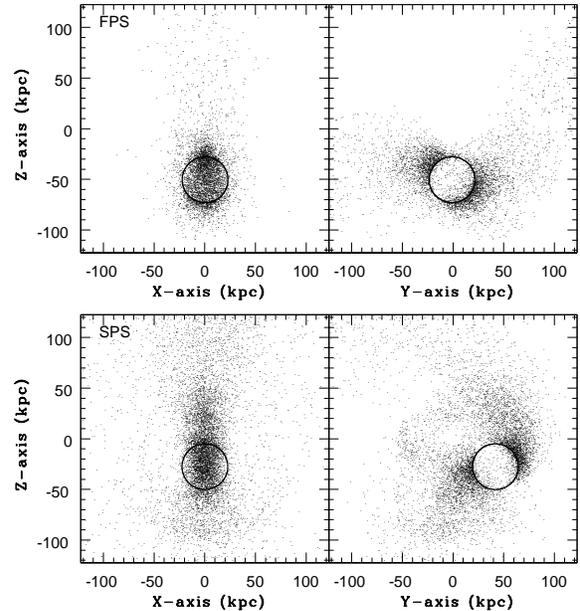,width=8.0cm}
\caption{
The distributions of the stripped LMC stellar halo particles projected onto
the $x$-$y$ plane (left) and the $y$-$z$ plane (right) for the FPS (upper two)
and the SPS (lower two). 
Only particles outside $R=3R_{\rm d, L}$ (where $R_{\rm d, L}$ is 
the initial stellar disk size and 7.5 kpc) are regarded as those stripped from
the initial LMC stellar halo. The thick solid line in each frame represents the
initial  halo size ($R_{\rm h, L}=3R_{\rm d, L}$).
}
\label{Figure. 3}
\end{figure}

\begin{figure}
\psfig{file=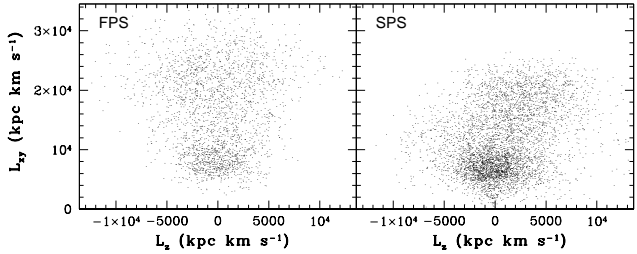,width=8.5cm}
\caption{
The distributions of the stripped LMC stellar halo particles on
the $L_{\rm z}-L_{\rm xy}$ plane  in the FPS (left) and 
the SPS (right).
Here  $L_{\rm z}$ ($L_{\rm x}$, $L_{\rm y}$) and $L_{\rm xy}$ are
the angular momentum component in the $z$ ($x$, $y$) direction
and ${(L_{\rm x}^2+L_{\rm y}^2)}^{1/2}$, respectively.
The stripped LMC stellar halo stars shown in Fig. 4 are plotted in this figure.
}
\label{Figure. 4}
\end{figure}

\begin{figure}
\psfig{file=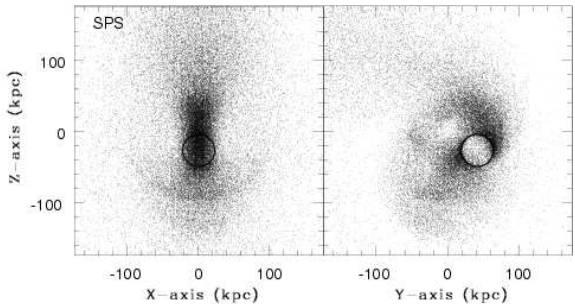,width=8.0cm}
\caption{
The distributions of the stripped LMC dark halo particles projected onto
the $x$-$y$ plane (left) and the $y$-$z$ plane (right) for the SPS.
Only particles outside $R=R_{\rm h, L}$
are  regarded as those stripped from
the initial LMC dark  halo. The thick solid line in each frame represents the
initial halo size ($R_{\rm h, L}=3R_{\rm d, L}$).
}
\label{Figure. 5}
\end{figure}

\begin{figure}
\psfig{file=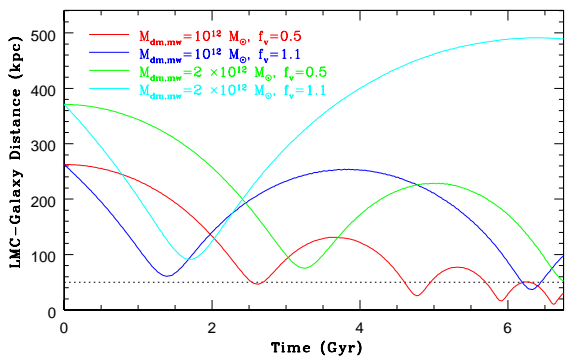,width=8.0cm}
\caption{
The same as Fig. 1 but for four models with different $f_{\rm v}$ and
$M_{\rm dm,mw}$: M1 (red), M4 (blue), M9 (green), and M11 (cyan).
The present LMC-Galaxy distance is shown by a black dotted line.
}
\label{Figure. 6}
\end{figure}

\section{Results}

\subsection{FPS versus SPS}

Fig. 2 shows how the LMC stellar halo is stripped by the strong tidal field
of the Galaxy to form a disk-like or  ring-like stellar structure with a long diffuse
stellar stream in the outer Galactic halo ($R<200$ kpc) in the standard SPS (M2).
After the  first pericenter passage ($T\sim 3$ Gyr), the outer part of the
LMC stellar halo is partially stripped to form diffuse leaning and trailing stellar
stream ($T=3.8$ Gyr). These stripped stars can be then trapped by the Galactic potential
to form a smaller ring-like structure with the radius  of $\sim 50$ kpc ($T=4.5$ Gyr),
while the LMC stellar halo continues to be stripped.
Finally an irregular ring-like stellar structure connected to the LMC and the outer
diffuse stellar stream can be formed after the second pericenter passage ($T=5.6$ Gyr). 
Owing to the initially extended structure of the LMC stellar halo,
the stripped stars do not form  narrow tidal streams.

Fig. 3 shows how the 3D distributions of the stripped LMC stellar halo particles
depend on the past orbits of the LMC.  Here stripped stars are those outside
the original stellar halo size ($R_{\rm h, L}=3R_{\rm d, L}=22.5$ kpc)
at $T=T_{\rm p}$.
Clearly, a thick disk structure along the polar axis (i.e., $z$-axis) can be 
seen for $|z|<50$ kpc in the SPS whereas such a disky structure can not 
be seen in the FPS. 
Furthermore, diffuse leading and trailing stellar streams can be seen 
in the $y$-$z$ projection for  the FPS
whereas a closed irregular ring can be seen in the SPS.
Since it is almost impossible for any models based on the FPS to show
a closed  ring just after the last pericenter passage,
the derived clear differences in distributions of stripped LMC halo stars
projected onto  the $y$-$z$ plane  between the two scenarios can be used
for distinguishing the two scenarios.

Fig. 4  demonstrates that the tidally stripped LMC stellar halo
can show a characteristic distribution
in  the  $L_{\rm z} - L_{\rm xy}$ plane,
where $L_{\rm z}$ ($L_{\rm x}$, $L_{\rm y}$) and $L_{\rm xy}$ are
the angular momentum component in the $z$ ($x$, $y$) direction
and ${(L_{\rm x}^2+L_{\rm y}^2)}^{1/2}$, respectively.
Distributions of halo stars on this  $L_{\rm z} - L_{\rm xy}$ plane
have been discussed in the context of the formation of the Galactic stellar halo
due to accretion of dwarf galaxies 
(e.g., Chiba \& Beers 2000). 
It should be stressed that only the stripped stars ($R>R_{\rm h, L}$)
shown in  Fig. 3 (i.e., not all stars) are plotted in this Fig. 4.
Owing to the diffuse and wide stellar distributions in the tidal streams
and the polar ring,  the stellar distribution in phase space does not 
show any strong concentration in the FPS and SPS.
However, there is a clear difference between the two in the sense that
the FPS shows a much larger number fraction of stars with 
$L_{\rm xy}$ larger than $2 \times 10^4$ kpc km s$^{-1}$ 
in comparison with the SPS.
The presence of stars with larger $L_{\rm xy}$ in the FPS is due to the fact
that the LMC is moving faster in the FPS.
Thus this appreciable difference in phase space can be used to distinguish 
the two scenarios.

The present study predicts that the stripped dark matter halo  of the LMC can
form a polar thick disk with a central hole (``dark polar ring'') 
in the SPS. 
A ``ripple'' (of ``shell'') structure can be clearly seen at $-100$kpc$<z<-40$kpc
and $-100$kpc$<x<100$kpc in the $x$-$z$ plane.
If dark matter particles outside $R>R_{\rm h, L}$ are regarded as those
consisting of the dark polar ring, 
then the total mass of the ring is $5.5 \times 10^{10} {\rm M}_{\odot}$
(i.e., about 61\% of the original dark halo mass) which corresponds
to  5.5\% of the total mass of the Galactic dark matter in the present model.
Although this dark polar ring would be 
currently impossible to be detected by observational studies,
it could dynamically influence orbital evolution of dwarf satellite galaxies
around the Galaxy, in particular, those in polar orbits (e.g., Carina dwarf galaxy).

\begin{figure}
\psfig{file=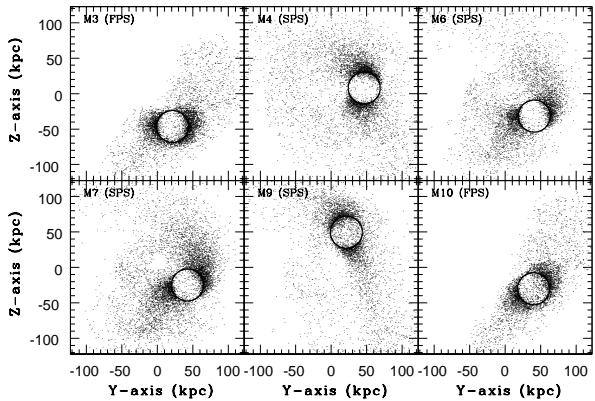,width=8.0cm}
\caption{
The same as Fig. 3 but for distributions of the stripped LMC halo
stars projected onto the $y$-$z$ plane in six different models,
M3, M4, M6, M7, M9, and M10. Whether each model is regarded as FPS or SPS
is indicated in the upper left corner of each frame (next to model number).
}
\label{Figure. 7}
\end{figure}

\begin{figure}
\psfig{file=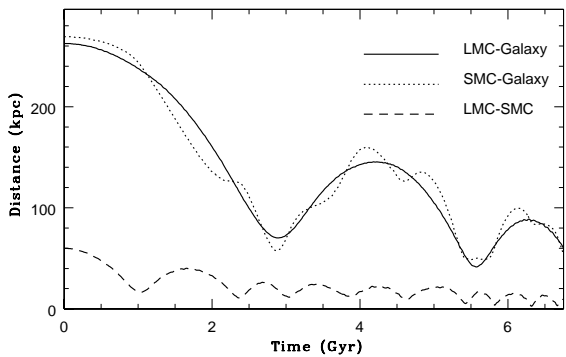,width=8.0cm}
\caption{
The time evolution of the LMC-Galaxy (solid), SMC-Galaxy (dotted), and LMC-SMC (dashed)
distances in the model M7 with SMC. 
}
\label{Figure. 8}
\end{figure}

\subsection{Dependence on model parameters}

\subsubsection{Orbit}

Fig. 6 shows the orbital evolution of the LMC in the four representative models
with different $f_{\rm v}$ and $M_{\rm dm, mw}$. Clearly, the LMC can not merge
within the last 6.8 Gyr in these four models, which demonstrates a possibility
that the LMC was accreted onto the Galaxy long time ago but has not merged with 
the Galaxy yet. 
Although the LMC initially has a larger total mass 
($M_{\rm t, L}=9.6 \times 10^{10} {\rm M}_{\odot}$),
it can lose a significant fraction of the mass owing to tidal stripping
by the Galaxy.  This mass loss can make the time scale of dynamical friction
significantly longer for the LMC.  
As a result of this,  the LMC can  not so quickly merge with the Galaxy
as it otherwise could if it keeps the large initial mass. 
The LMC in the model M4 with $f_{\rm v}=1.1$
can finally show $R_{\rm p} \sim 50$ kpc corresponding
to the present distance of the LMC from the center of the Galaxy.
The LMC can not merge with the Galaxy within $\sim 10$ Gyr in this model.

Given that the adopted $f_{\rm v}=1.1$ in the model M4 is predicted
to be typical in recent $\Lambda$CDM cosmological simulations (e.g., Wetzel 2011),
this means that the LMC orbit needs  not be unusual one in a $\Lambda$CDM universe.
Only the model with lower $f_{\rm v}$ for  $M_{\rm mw, dm}=2 \times 10^{12} {\rm M}_{\odot}$
can show $R_{\rm p} \sim 50$ kpc within the last 6 Gyr. Such a lower $f_{\rm v}$ is
not typical in the $\Lambda$CDM simulations, which implies that the LMC could be
an unusual object if the Galaxy is as massive as 
$M_{\rm mw, dm}=2 \times 10^{12} {\rm M}_{\odot}$.

Fig. 7 shows the distributions of the stripped LMC stellar halo particles around the Galaxy
projected onto the $y$-$z$ plane for six representative models. 
Since the distributions of the stars projected onto the $x-z$ plane 
are less significantly different  between FPS and SPS,
we here focus on the distributions projected onto the $y$-$z$ plane.
Clearly,  the model M3, which is regarded as FPS, does not show a  ring-like 
distribution for the stripped LMC stellar halo stars. 
This means that irrespective of orbital models,
FPS can not show a ring-like distribution of the stripped stars along the polar axis
of the Galaxy. 
The model M9 (which is regarded as SPS) 
shows a ring-like distribution of the stripped stars whereas
the model M10 (FPS) does not. 
This clear difference between FPS and SPS in models with larger masses of the Galaxy
($M_{\rm mw, dm}=2 \times 10^{12} {\rm M}_{\odot}$) strongly suggests that
the stripped LMC stellar halo stars can be used for distinguishing FPS and SPS.
 
\subsubsection{$R_{\rm h, L}$}

The model M6 has a smaller size of the LMC stellar halo ($2R_{\rm h, L}=15$ kpc) so that
the halo can not be so strongly influenced by the tidal field of the Galaxy in comparison
with the model M2. Fig. 7 shows that there  is no significant difference in the distribution
of the stripped LMC stellar halo stars between M2 and M6 (i.e., the two models with
the same orbit of the LMC). This result combined with that derived for
the model M2 suggests  that as long as the LMC has an extended
stellar halo ($R_{\rm h, L} \ge 2R_{\rm d, L}$), the stripped halo stars can have 
fossil record of the past orbit of the LMC.
If the LMC stellar halo has $R_{\rm h, L} \le  R_{\rm d, L}$, then
the stripping of the halo stars become less efficient. However,
such models with $R_{\rm h, L} \le  R_{\rm d, L}$ are inconsistent
with observations of galactic stellar halos by Zibetti et al. (2004).

\subsubsection{SMC}

In order to discuss how the LMC-SMC interaction can influence the final distribution
of the stripped LMC halo stars in the outer halo of the Galaxy,  we construct a N-body
model for the SMC. The SMC has an initial total mass of $4.8 \times 10^9 {\rm M}_{\odot}$
and is modeled as a bulge-less disk galaxy embedded in a massive dark matter halo.
The structural and kinematical properties of stellar disks and dark matter halos
are assumed to be self-similar between the LMC and the SMC. The SMC has an exponential
stellar disk with the mass of $2.8 \times 10^8 {\rm M}_{\odot}$ and
the size of 2.0 kpc. The initial distance of the SMC from the center of the LMC 
is a free parameter and denoted as $R_{\rm i, S}$. The initial velocity of the SMC 
with respect to that of the LMC is given as $0.6v_{\rm cir, L}$, where $v_{\rm cir, L}$
is the circular velocity of the LMC at $R_{\rm i, S}$.
 The orbital evolution of the SMC depends not only
on the orbital parameters of the LMC but also on $R_{\rm i, S}$. We here describe the
results for the models M7 ($R_{\rm i, S}=8R_{\rm d, L}$), 
because the LMC and the SMC do not merge so quickly with each other in this model.
Although we  have investigated
a number of models with different $R_{\rm i, S}$,
the models with smaller $R_{\rm i, S}$
like M8 with $R_{\rm i, S}=6R_{\rm i, S}$ show a short time scale of LMC-SMC merging
(within $\sim 4$  Gyr). They are not so useful for discussing physical properties
of the present LMC-SMC system which has not completely merged yet.

Fig. 8 shows the orbital evolution of the SMC with respect to the Galaxy and the LMC
in the model M7. This is the first orbital model of the SMC based on a fully self-consistent
N-body simulation in which all of the three galaxies are represented by N-body particles.
Dynamical friction between the Galaxy, the LMC, and the SMC is self-consistently
included so that the orbit of the SMC can be more precisely predicted in this model.
Clearly, the Clouds can keep their binary status for the last 5 Gyr, if they were initially
strongly bound.  Owing to dynamical friction between the LMC and the SMC,
the Clouds appear to be almost merged at $T \sim6$ Gyr. Fig. 7 shows that the stripped
LMC halo stars in this model has a ring-like distribution along the polar axis of the 
Galaxy and the distribution is quite similar to that in the model M2. Thus the SMC
can not change significantly the spatial distribution of the stripped LMC halo stars
in the outer Galactic halo owing to its low mass.

\section{Discussion}

\subsection{Relation to cosmological N-body simulations}

Recent theoretical studies based on high-resolution collisionless N-body simulations
have investigated orbital properties (e.g., eccentricities and pericenter distances)
of satellite galaxies around luminous ones like the Galaxy 
(e.g., Benson 2005; Lux et al. 2010; Boylan-Kolchin et al. 2011, B11; Wetzel 2011). 
One of their results relevant to the present study is distributions of 
normalized total velocities ($V_{\rm tot}$  in 
Fig.2  by  Wetzel 2011) which correspond to $f_{\rm v}$ in the present study.
The predicted $f_{\rm v}$ distribution in Wetzel (2011)
has a peak value of 1.15 and shows only a 
small fraction of satellites with $f_{\rm v} \le 0.6$.
This means that the orbits used in models with low $f_{\rm v}$ $\le 0.6$
are  less common
in the $\Lambda$CDM model: the present models with $f_{\rm v}\sim 1$ 
would be more appropriate for describing the orbit of the LMC.
Given that both low and high $f_{\rm v}$ models show clear differences 
in 3D distributions of stripped LMC halo stars between 
FPS and SPS,  the above less consistent models do not become a problem in 
the present study.

The LMC in the models with $f_{\rm v} \sim 1$ and $R_{\rm i} \sim 260$ kpc can finally have 
$R_{\rm p} \sim 50$ kpc (at $T \sim 6$ Gyr)
owing to the orbital decay of the LMC caused by dynamical friction against the Galaxy.
Given that $f_{\rm v} \sim 1$ is typical in recent cosmological simulations
(Wetzel 2011),  
this result means that if the LMC has a typical orbit for  satellites  in the Galaxy,
then it should have been accreted onto the Galaxy more than $\sim 6$ Gyr ago.
This also suggests that successful formation models of MS and LAs
will be able to be constructed by using results of cosmological N-body simulations
on orbital properties of LMC-type satellites in Galaxy-type halos.
The LMC even in the models with lower $f_{\rm v}$ ($\sim 0.5$) can not merge
with the Galaxy within the last 6 Gyr.  This demonstrates a possibility  that
the LMC was accreted onto the Galaxy more than $\sim 6$ Gyr ago and has been
interacting with the Galaxy since then.

It is well known that the observed locations, kinematics, fine-structures
(e.g., bifurcation) in MS and LAs can give very
strong constraints on the orbital evolution of the LMC and the SMC
for a given mass model of the Galaxy
(e.g., GN96; YN03; C06; DB11).
Previous models that reproduced very well the observed bifurcated MS and long
LA predict that the LMC and SMC have passed their pericenter 
about  the Galaxy at least twice for the 
last 2$-$3  Gyr (C06; DB11).
Previous dynamical models for the formation of MS based on FPS have reproduced
neither the observed location of MS (Fig. 21 in Bekki \& Chiba 2009) nor the bifurcated
structure of MS (Besla et al. 2010). These models also failed to reproduce the
observed properties of the LAs so far: their models can not be regarded as successful
MS formation models.
These results imply  that the LMC arrived within the virial radius of the Galaxy (for
the first time) more than 4$-$5 Gyr ago.
The question is whether this earlier arrival of the LMC is consistent with
the latest results of dissipationless cosmological N-body simulations on
accretion processes of LMC-type satellite galaxies onto Galaxy-type halos.

Using the results of the Millennium II simulation,
B11 investigated the epochs ($t_{\rm fc}$) of first accretion 
onto Galaxy-mass galaxies and orbital properties for the ``LMC analogs''
that have masses ranging from $8 \times 10^8 {\rm M}_{\odot}$
to  $3.2 \times 10^{11} {\rm M}_{\odot}$ at the accretion epochs.
They showed that (i)  only $\sim$30\% of LMC analogs  have $t_{\rm fc} \le 2$ Gyr 
(their Fig.1) ,
(ii) LMC analogs with $t_{\rm fc} >4$ Gyr have lower angular momentum
(Fig. 2),
(iii) LMC analogs with $t_{\rm fc} >4$ Gyr are more likely to be strongly bound 
(Figs. 3 and 4),
and (iv) about 50\% of LMC analogs with $t_{\rm fc}>4$ Gyr have orbital eccentricities ($e$)
 smaller than 0.6
(Fig. 5).
These results  are consistent  with the previous successful MS models (e.g., GN96;
YN03, and  DB11) that 
predict bound orbit of the LMC with $e \sim 0.4$
and  at least two  pericenter passages of the LMC in the Galaxy  
(i.e., $t_{\rm fc}>4$ Gyr).
If we estimate the specific angular momentum of the LMC ({\it \~j})
in GN96 by adopting the same method as used in B11,
then {\it \~j} is 0.26 for $r_{\rm vir, mw}= 245$ kpc.
The results in Fig. 2 in B11 show that about 60\%
of early-accreted LMC analogs with $t_{\rm fc}>4$ Gyr
have  {\it \~j }$< 0.26$. This means that the predicted orbit of the LMC in GN96
is a common one in B11.
We thus conclude that the orbital properties
in the above successful MS formation models based on the early accretion
of the LMC are consistent with those 
of LMC analogs predicted from  recent cosmological simulations
by B11.

Our conclusion above therefore appears to be at odd with B11 which
concluded that the LMC was accreted within the past four Gyr
and is currently making its first pericenter passage about the Galaxy.
Clearly, their Fig. 4 shows that about 60\% of LMC analogs are accreted
onto the Galaxy more than 4 Gyr ago (i.e., $t_{\rm fc}>4$ Gyr). This result
appears to be inconsistent with the above conclusion by B11.
This apparent 
inconsistency is  due to the fact that B11 used only observations
by K06 and thereby estimated orbital energy and angular momentum of the LMC in order to
find LMC analogs that have orbital properties similar to the observed ones: their conclusions
depend on the adopted assumptions on the orbital angular momentum and energy of
the LMC.
Now new observational results on the proper motion of the LMC 
(e.g., C09;  V10) have shown that 
the orbital energy and angular momentum of the LMC are  significantly different
from those by K06.
These significant differences between different observations
imply that it would be worthwhile for cosmological N-body simulations 
to find the best LMC analogs separately for each of  observations 
by K06,  C09,  and V10.

\subsection{Fossil records in the outer Galactic halo}

The present study has first shown that the outer regions  of the Galactic stellar
halo ($R>50$ kpc), in particular, those along the Galactic polar axis,
can have fossil information on the past 3D orbit of the LMC.
If the LMC has already experienced the pericenter passage at least two times,
then the stars stripped from the LMC stellar halo can have a thick disk
or ring-like  distribution along the $z$-axis in the Galactic coordinate.
This  polar disk (or ring) has a mass of 
$\sim 10^7 {\rm M}_{\odot}$ ($\approx 0.2 M_{\rm h, L}$ ) 
and thus can consist of
$\sim 0.5$\% of the entire Galactic stellar halo mass. The rotating kinematics of the
polar disk and the spin vector of the polar disk (i.e., the $x$-direction)
would be distinguished from those of  other possible stellar substructures formed from
tidal destruction of dwarf galaxies
in the Galactic stellar halo. 
Therefore, if the polar disk really exists in the
outer stellar halo,  then it would be feasible for ongoing and future observations
to detect them along the Galactic polar axis.

Previous models showed that 
the MS along the polar axis of the Galaxy
can be formed from efficient  stripping of gas from the SMC due to 
tidal interaction between the LMC, SMC, and the Galaxy (YN03).
Their models also predicted that  there should be no/little stars
along the MS, because tidal stripping is efficient only for gas that is 
initially located
in the outer part of the SMC in comparison with stars (See Fig. 8 in YN03).
Therefore, if observations discover  stars in the MS region,
then the stars are unlikely to 
originate from  the stellar disk of the SMC.
However, if the SMC also has an extended stellar halo, as suggested by recent observations
by No\"el \& Gallart (2007), then the stellar halo can be stripped by the Galaxy
during the LMC-SMC-Galaxy tidal interaction to form a stellar stream/disk
along the MS. 
If this is the case, the stellar structure along the MS would consist of stars from 
metal-poor stellar
halos of the LMC and the SMC.

Guhathakurta \& Reitzel (1998) searched for stars in the $5^{'}\times7^{'}$ field 
of the MS IV region but could not find possible evidence for the present of stars
in the region.  They therefore suggested that the total mass ratio of stars 
and neutral hydrogen  gas ($M_{\ast}/M_{\rm HI}$) is less than 0.1.
Given that the mass of the MS is $\sim 5 \times 10^8 {\rm M}_{\odot}$ (e.g., 
Br\"uns et al. 2005),  $M_{\ast}/M_{\rm HI}$, which is the mass ratio
of the stripped LMC stellar halo to the MS here,   can be well less than 0.05. Therefore,
the presence of the polar disk/ring composed of the stripped LMC halo stars
would not be inconsistent with the above observation. 
The narrow observational field in Guhathakurta \& Reitzel (1998)
would possibly have prevented
them to detect the high-density regions of the possible  
polar disk/ring.
Br\"uck and Hawkins (1983) did not find stars brighter than 20.5 mag in the B-band
for the MS
and thus suggested that the MS is unlikely to contain intermediate-age stars
($\sim 3 \times 10^9$ yr). 
Given that this observation was difficult to detect old faint stars,
the non-detection of stars does not necessarily mean
the absence of any stars in the MS.

Thus, although  observational studies have  not revealed evidence for the presence of
old stars in the MS so far,
this non-detection could be due largely to observational constraints (e.g.,
narrow field and shallow photometric survey etc).
The present study predicts that the MS and the possible polar disk/ring
can overlap to some extent  in the projected distributions on the sky,
their distributions in the 3D space are yet quite different.
Thus a  very wide and deep photometric survey {\it along and nearby the MS}
is necessary to map the entire stellar distribution of 
the possible stellar polar disk/ring. 
Future observational surveys such as the Southern Sky Survey by the SkyMapper
telescope (Keller et al. 2007) would be ideal for the detection of the polar disk/ring.

It should be stressed that the present results depend strongly on
whether the LMC  has an extended stellar halo well before
its commencement of tidal interaction with the Galaxy: if the LMC initially
does not have the halo, then the outer stellar halo of the Galaxy would not have
valuable information on the past 3D orbit of the LMC.
Extensive observational studies on physical properties
of the outer part of the LMC based on photometric properties of stars
(e.g., color-magnitude diagrams)
are currently ongoing (e.g., Saha et al. 2010).
The physical extension and projected 2D distribution of the possible LMC stellar halo
revealed by these studies will assess the viability of the present new idea
to give constraints on the past orbit of the LMC.

\subsection{Other possible evidence for early accretion of the LMC ?}

Using the results of numerical simulations on dynamical properties of substructures
in a Milky Way-like halo in a $\Lambda$CDM model,
Li \& Helmi (2008) showed that accretion of  a small group of galaxies  
on to the Galaxy about 8 Gyr ago can explain the polar distribution of satellite
galaxies in the Galaxy.  Bekki (2008) suggested that if the LMC was a central galaxy
of a group being accreted onto the 
Galaxy  (i.e., with other small galaxies, like the SMC, Carina, and Fornax),
then the LMC now can have a ``common  halo''  that surrounds both the LMC
and the SMC.If this is the case, the stripped low-mass dwarfs from the LMC
group might well have a
polar distribution like the stripped LMC stellar halo.
The observed thin disky distribution  of the Galactic satellite
galaxies along the Galactic polar axis (e.g., Metz et al. 2009)
could  be thus possible evidence of early accretion of the LMC onto the Galaxy,
if the LMC was accreted onto the Galaxy as a group.

Yoon \& Lee (2002) revealed an intriguing  distribution of seven  metal-poor ([Fe/H]$<-2$) 
globular clusters (GCs) along the polar-axis of the Galaxy and suggested that
the GC distribution is consistent with  accretion of a satellite galaxy which
lost their GCs by tidal stripping of the Galaxy. The GCs along the polar-axis
have ring-like or disky spatial distribution and rotational kinematics,
which appear to be similar to dynamical properties of the stripped LMC stellar halo stars 
in SPS of the present study.
These results imply that some of
original GCs in the LMC  halo were tidally stripped during 
the accretion of the LMC more than $\sim 4$ Gyr ago to become the  halo GCs with
a disky spatial distribution along the polar axis of the Galaxy.
This picture would be  consistent with the observed similarities in physical properties
of GCs (e.g., locations on [Fe/H]-${\rm P}_{\rm ab}$ relation, where ${\rm P}_{\rm ab}$
is the mean period of type ab RR Lyraes) between the seven GCs and the LMC GCs.

\section{Conclusions}

We have investigated dynamical evolution of the stellar halo of the LMC orbiting
the Galaxy using fully self-consistent N-body simulations in which the LMC, the SMC,
and the Galaxy are represented by collisionless N-body  particles. We also have investigated
orbital evolution of the LMC for different initial velocities of the LMC at its
first passage of the virial radius of the Galaxy.
This is for the purpose of discussing  whether
orbital properties of the LMC in
the previous successful formation models for MS and LAs are consistent with
those predicted from recent cosmological simulations based on a $\Lambda$CDM model.
The main results are summarized as follows.

(1) The tidal interaction between the LMC and the Galaxy can strip stars initially in the
extended stellar halo of the LMC and the stripping
processes depend strongly on the past orbit of the LMC.
For example,
if the epoch of the first LMC accretion onto the Galaxy
from outside its viriral radius is more than $\sim 4$  Gyr ago
(i.e., at least two pericenter passages),
the stars stripped from the stellar halo of the LMC can form an irregular
polar ring or a thick  disk
with a size of $\sim 100$ kpc and rotational kinematics.
On the other hand, if the LMC  was first accreted onto
the Galaxy quite recently ($\sim 2$ Gyr
ago),
the stripped stars form shorter leading and trailing stellar stream at $R=50-120$ kpc.
The distributions of the  stripped stars
in phase space between the two cases can be significantly different.

(2) Differences in dynamical properties of the stripped LMC halo stars
between FPS and SPS
do not depend on model parameters such  as $M_{\rm dm, mw}$, $R_{\rm h, L}$, and $f_{\rm v}$.
Also the SMC is found to be unable to significantly change the final distribution
of the stripped LMC stellar halo around the Galaxy.
Therefore the derived differences between FPS and SPS
suggest that if we compare the observed three-dimensional (3D)  distribution
and kinematics of the
outer Galactic stellar halo along the polar-axis,
then we can give strong constraints on the past orbit of the LMC.
Future observational surveys such as the Southern Sky Survey by the SkyMapper
telescope  would be ideal for the detection of the polar stellar disk/ring
possibly formed from the past LMC-Galaxy interaction.
The Galaxy can have a massive polar dark matter ring formed from
the LMC-Galaxy interaction, if the LMC was accreted onto the Galaxy more than several
Gyr ago. The observed disky distributions of the Galactic satellite galaxies
and  seven Galactic halo GCs along the polar axis of the Galaxy
could be possible evidence of the early accretion of the LMC.

(3) The LMC can lose its mass, energy, and orbital angular momentum during its orbital
decay caused by dynamical friction against the dark matter halo of the Galaxy. 
The significant  mass loss of the LMC during its orbital evolution can 
make the timescale of dynamical friction significantly longer.
As a result of this,
the LMC can not merge with the Galaxy  within $\sim$  6 Gyr after its first passage of
the Galactic virial radius, even if the initial total mass of the LMC 
is about 10\% of that of the Galaxy. 
The LMC with $f_{\rm v} \sim 1$ ($f_{\rm v,y}=0.5$ and $f_{\rm v,z}=0.9$)
that is predicted to be typical
in recent $\Lambda$CDM cosmological
simulations can not merge within $\sim 10$ Gyr for 
$M_{\rm t, L}=9.6 \times 10^{10} {\rm M}_{\odot}$ and
$M_{\rm dm, mw}=10^{12} M_{\odot}$.
These results allow  the previous
successful MS formation models
which assume  that the LMC and the Galaxy have been dynamically interacting with
each other without merging at least a few Gyr.

(4) The LMC in the models with $f_{\rm v} \sim 1$ 
can finally have $R_{\rm p}$ ($\sim 50$ kpc) and 
$V_{\rm p}$ ($300-380$ km s$^{-1}$)  that are observed in the
present LMC.  This result implies that the orbit of the LMC 
needs  not be unusual one in the $\Lambda$CDM model.
This also suggests that it is promising for future formation models of MS and LAs
based on cosmological simulations to reproduce both the present location/velocity
of the LMC and the physical properties of MS and LAs in a self-consistent manner.

(5) The previous successful formation models of MS and LAs predict at least
two pericenter passages of the LMC about the Galaxy and thus accretion of the LMC
onto the Galaxy (i.e., the first passage of the Galactic
virial radius) more than 4 Gyr ago.
The models also  predict  $e \sim 0.4$, {\it \~j}$\sim 0.3$, and strongly bound
orbit for the LMC.  
Orbital properties of
early-accreted LMC analogs ($t_{\rm fc}>4$ Gyr) 
predicted  from recent cosmological simulations 
are therefore consistent with those in the successful formation models of MS and LAs. 
This consistency strongly suggests that the LMC was accreted onto the Galaxy
more than 4 Gyr ago so that the bifurcated MS and 
elongated LAs could be  formed as a result of tidal interaction
between the LMC, the SMC, and the Galaxy.

We thus conclude that the LMC was accreted onto the Galaxy more than 4 Gyr ago
(i.e., the first pericenter passage more than $2-3$ Gyr ago).
This early accretion 
suggests that the global star formation 
of the LMC could have been enhanced by its dynamical interaction
(Bekki \& Chiba 2005) and hydrodyanmical one (Mastropietro et al. 2009)
with the Galaxy since more than $2-3$ Gyr ago (i.e., after the first pericenter
passage).   This LMC-Galaxy interaction which can trigger star formation
can provide physical explanations
both for the observed enhancement of star formation in the LMC about a few Gyr ago
(e.g., Gallagher et al. 1996) and for the unusually blue colors of the LMC as a massive
satellite in Galaxy-scale halos (Tollerud et al. 2011).
Furthermore, the long-term LMC-SMC-Galaxy interaction, which is possible in
early accretion of the LMC,
can repetitively enhance star formation in the SMC for the last few Gyr (YN03), 
which is indeed  observed (e.g., Harris \& Zaritsky 2004). 
Thus, the observed star formation histories of the LMC and the SMC,
bifurcated MS,  and elongated LAs
strongly suggest that 
the LMC first passed the Galactic virial radius more than 4 Gyr ago.

\section{Acknowledgment}
I am  grateful to the anonymous referee for constructive and
useful comments that improved this paper.
KB acknowledge the financial support of the Australian Research Council
throughout the course of this work.
The work was supported by iVEC through the use of advanced computing resources located at the University of Western Australia.

\end{document}